\documentclass[sigconf]{acmart}
\usepackage{adjustbox}
\usepackage{multirow}
\usepackage{array}
\usepackage{tabu}
\usepackage{lipsum}
\usepackage{booktabs}
\usepackage{graphicx}
\usepackage{csquotes}
\usepackage{pdfpages}
\usepackage{pifont}
\usepackage{color}
\usepackage{colortbl}
\usepackage{alphabeta}
\usepackage{threeparttable}

\definecolor{LightGray}{gray}{0.95}

\newcolumntype{P}[1]{>{\centering\arraybackslash}p{#1}}
\newcolumntype{L}[1]{>{\raggedright}p{#1}} 
\newcolumntype{R}[1]{>{\raggedleft}p{#1}}  

\newcommand{\xmark}{\ding{55}}%

\begin{document}

\title{Why Do Developers Get Password Storage Wrong? \\ A Qualitative Usability Study}

\author{Alena Naiakshina}
\authornote{These authors contributed equally to this work.}
\affiliation{
  \institution{University of Bonn}
}
\email{naiakshi@cs.uni-bonn.de}

\author{Anastasia Danilova}
\authornotemark[1]
\affiliation{
  \institution{University of Bonn}
}
\email{danilova@cs.uni-bonn.de}

\author{Christian Tiefenau}
\affiliation{
  \institution{University of Bonn}
}
\email{tiefenau@cs.uni-bonn.de}
\author{Marco Herzog}
\affiliation{
  \institution{University of Bonn}
}
\email{herzog@cs.uni-bonn.de}
\author{Sergej Dechand}
\affiliation{
  \institution{University of Bonn}
}
\email{dechand@cs.uni-bonn.de}
\author{Matthew Smith}
\affiliation{
  \institution{University of Bonn}
}
\email{smith@cs.uni-bonn.de}

\renewcommand{\shortauthors}{}

\begin{abstract}
Passwords are still a mainstay of various security systems, as well as the cause of many usability issues. For end-users, many of these issues have been studied extensively, highlighting problems and informing design decisions for better policies and motivating research into alternatives. However, end-users are not the only ones who have usability problems with passwords! Developers who are tasked with writing the code by which passwords are stored must do so securely. Yet history has shown that this complex task often fails due to human error with catastrophic results. While an end-user who selects a bad password can have dire consequences, the consequences of a developer who forgets to hash and salt a password database can lead to far larger problems. In this paper we present a first qualitative usability study with 20 computer science students to discover how developers deal with password storage and to inform research into aiding developers in the creation of secure password systems. 

\end{abstract}




\copyrightyear{2017}
\acmYear{2017}
\setcopyright{acmlicensed}
\acmConference{CCS '17}{October 30-November 3, 2017}{Dallas, TX, USA}\acmPrice{15.00}\acmDOI{10.1145/3133956.3134082}
\acmISBN{978-1-4503-4946-8/17/10}
\maketitle
\section{Introduction}

In the last decade usable security research has focused mainly on issues faced by end-users. In their paper ``Developers are not the enemy'', Green and Smith \cite{green2016developers} argue that usable security research needs to be extended to study and help developers who have to interact with complex security APIs. 

Storing and authenticating user login data is one of the most common tasks for software developers~\cite{nadi2016jumping}. At the same time, this task is prone to security issues~\cite{green2016developers}. Frequent compromises of password databases highlight that developers often do not store passwords securely - quite often storing them in plain text ~\cite{florencio2014administrator, vance2010if, finkle2012linkedin, kamp2012linkedin}. 

To gain an understanding on where and why developers struggle to store passwords securely, we conducted the first usability study of developer behavior when tasked with writing code to store passwords and authenticate users. 
Since this is the first work in this domain, we chose to conduct a qualitative study with the ability to conduct in-depth interviews to get feedback from developers.

We were interested in exploring two  particular aspects: Firstly, do developers get things wrong because they do not think about security and thus do not include security features (but could if they wanted to)? Or do they write insecure code because the complexity of the task is too great for them? Secondly, a common suggestion to increase security is to offer secure defaults. This is echoed by Green and Smith \cite{green2016developers} who call for secure defaults for crypto-APIs. Based on this suggestion, we wanted to explore how developers use and perceive frameworks that attempt to take the burden off developers. 
 
To offer insights into these questions, we conducted three pilot studies and a final 8-hour qualitative development study with 20 computer science students. The use of computer science students for software engineering studies is considered an acceptable proxy for developers \cite{host2000using, berander2004using,Svahnberg:2008cj, salman2015students}. The students were given a role-playing scenario in which they were tasked to include registration and authentication code into an existing stub of a social networking application. To gather insights into task framing, half of the students were told to store the passwords securely and the other half were not. This latter group of students would have to decide on their own volition whether, since they were writing code for storing user passwords, they should store them securely. Also, one half of each group was given the Spring Framework to work with, which contains helper classes for password storage, while the other half had to use the Java Server Faces (JSF) Framework in which they would have to implement everything themselves. 

Our key insights are as follows: 
\begin{itemize}
    \item \textbf{Security knowledge does not guarantee secure software:} Just because participants had a good understanding of security practices did not mean that they applied this knowledge to the task. Conversely, we had participants with very little security knowledge who created secure implementations. 
     \item \textbf{More usable APIs are not enough}: Even when APIs offer better support for developers with ready-to-use methods which implement secure storage correctly, this is not enough if these methods are opt-in, as they were with all password storage APIs we found.
    \item \textbf{Explicitly requesting security is necessary}: Despite the fact that storage of passwords should self-evidently be a security sensitive task, participants who were not explicitly told to employ secure storage stored the passwords in plain text.
    \item \textbf{Functionality first, security second:} Very similarly to end-users, most of our participants focused on the functionality and only added security as an afterthought - even those who were primed for security.
    \item \textbf{Continuous learning:} Even participants who attempted to store passwords securely often did so insecurely because the methods they learnt are now outdated. Despite knowing that security is a fast moving field, participants did not update their knowledge.
    \item \textbf{Conflicting advice}: Different standards and security recommendations make it difficult for developers to decide what is the right course of action, which creates frustration. 

\end{itemize}
Since this is also one of the first studies of security APIs, we also gained several insights into running such studies which might be helpful for the community. 
\begin{itemize}
    \item \textbf{Think Aloud:} While Think Aloud is considered a good method to gather insights into study participants' mental process in end-user studies, we did not find it as useful for developer studies. Both the length and the complexity of the tasks seem to be a hindrance. In addition, evaluating Think-Aloud data is very work intensive.
    \item \textbf{Task framing:} Developer studies need to take great care when framing their tasks. Giving developers very small focused tasks concerning crypto libraries will not uncover most of the problems we saw in the more realistic context of complex tasks, which mix security and functionality development. Both types of studies are worthwhile; however, the decision needs to be taken consciously according to the goal of the studies.
    \item \textbf{Qualitative research reveals essential insights:} Prior research often focused on only examining the end-solution and analyzing survey responses in developer studies. We found that qualitative investigations can dig deeper and reveal misconceptions and lack of knowledge even if the solutions are secure. 
\end{itemize}

\section{Related Work}
\subsection*{Passwords}
There is a huge body of usable security research into passwords
~\cite{shay2010encountering, komanduri2011passwords, mazurek2013measuring, bonneau2012science, weir2010testing, pwdperceptions:chi16, passwordpolicies:tissec16, passwordcreation:soups15, longpasswords:chi14, ur2012helping, meters:chi17, egelman2013does}.
Comparatively few studies have been done analyzing how developers cope with end-user password storage.

Bonneau and Preibusch~\cite{bonneau2010password} carried out an empirical analysis of password implementations deployed on the Internet. They evaluated 150 websites which offered free user accounts and relied on user-chosen textual passwords for authentication. All of the sites showed inconsistent implementation choices impacting security. Bonneau and Preibusch found evidence that websites storing payment details as well as user data provided appropriate password security. 
By contrast, the worst security practices were found on content websites having few security incentives. Most of them suffered from a lack of encryption to protect transmitted passwords. Further, storage of cleartext passwords in server databases with little protection of passwords against brute force attacks were commonly-applied practices. Bonneau and Preibusch concluded that these sites deploy passwords primarily for psychological reasons, both as a justification for collecting marketing data and as a way to build trusted relationships with customers.

Acar et al. \cite{wermke2017security} conducted an experiment with active GitHub users in order to explore the validity of recruiting convenience samples as representatives for professional developers in security related developer studies.
One task covered secure storage of user credentials. A number of participants stored the passwords in plain text or used unsafe hashing functions or static salts.
Neither the self-reported status as student or professional developer, nor the security background of the participants correlated with the functionality or security of their solutions.

Prechelt~\cite{prechelt2011plat_forms} performed an exploratory study comparing Web development platforms (Java EE, Perl, and PHP). 
Nine teams comprising professional developers had 30 hours to implement a Web-based application. The resulting applications were analyzed with regard to several aspects, such as usability, functionality, reliability, security, structure, etc. 
The findings suggested that wide differences in platform characteristics were presented in some dimensions, although possibly not in others. 

Finifter and Wagner~\cite{finifter2011exploring} conducted a further evaluation of the code produced in the competition of Prechelt. They searched for correlations between programming languages/framework support for security aspects and the number of vulnerabilities. They did not find a relationship between a programming language and application security. However, automatic features were effective at preventing vulnerabilities in general, but not for secure password storage. Although many of the used frameworks provided ``automatic'' support for storing end-users' passwords securely, the developers did not employ them. These findings motivated us to conduct a qualitative study to gain deeper insights into why developers do not store passwords securely.  

Bau et al.~\cite{bau2012vulnerability} explored Web application vulnerabilities. They used a metric to search for correlations between the vulnerability rate of a Web application and (1) the nature of a developer team (startup company vs. freelancers), (2) security background knowledge of developers, and (3) programming languages. They found that Web applications implemented with PHP or by freelancers showed higher rates of vulnerabilities. Regarding password storage, they observed significant discrepancies between the expertise and the resulting implementation solutions of freelancers. They concluded that further research is required to better understand developers and their practices. As with the previous study, this study only looked at the code and said nothing about the thoughts of the developers.

\subsection*{Usability of Crypto APIs}

In order to protect sensitive data processed by current applications, developers need to use cryptographic APIs. However, due to their complexity, developers often struggle to use them correctly. 

Egele et al.~\cite{egele2013empirical} investigated the misuse of cryptographic APIs for Android application developers. Of the more than $11\,000$ Android apps analyzed, 88\% featured at least one of six cryptographic errors.
Further, Lazar et al.~\cite{lazar2014does} found that 80\% of the analyzed Common Vulnerabilities and Exposures (CVEs) resulted from misuse of cryptographic APIs by developers.

Nadi et al.~\cite{nadi2016jumping} investigated the problems developers face when using cryptography APIs. They analyzed the top 100 Stack Overflow questions regarding cryptography questions and afterwards conducted a study with 11 developers recruited from Stack Overflow. They also analyzed 100 randomly selected public GitHub repositories using Java cryptography APIs and surveyed 37 developers. They identified poor documentation, lack of cryptography knowledge, and bad API design as the main obstacles developers struggle against. 

Stylos and Myers~\cite{stylos2008implications} conducted studies with developers to examine API design usability with regard to method placement. They created two different versions of three different APIs and asked ten developers to implement three small tasks. 
They found that developers solved a task faster when using APIs in which the classes from which they started their exploration included references to the other classes they needed. 

Green and Smith ~\cite{green2016developers} discussed the issue of security API usability providing examples and gave high-level recommendations to reduce developer errors. Gorski et al. ~\cite{gorski2016towards} analyzed studies investigating API usability and extracted usability characteristics relevant for security API usability.

Fahl et al.~\cite{fahl2013rethinking} conducted qualitative interviews with software developers who deployed malfunctioning SSL issues revealing usability problems in the API design. They presented a framework in order to help developers implement secure software.

Security behavior in general was investigated by Redmiles et al.~\cite{redmiles2016think,redmiles2016learned}. They revealed that people with IT skills accept security advice from different sources as well as reject advice for different reasons, such as handing responsibility to other actors.

Acar et al. ~\cite{acar2016you} examined whether different documentation resources influence the security of programmers' code. They conducted a between-subjects laboratory study, where developers were provided with a skeleton Android app and asked to complete four programming tasks based on the skeleton: the storage of data, the use of HTTPS, the use of ICC, and the use of permissions. Their main finding was that developers who were assigned to use Stack Overflow created less secure code.

Acar et al. \cite{acar2017comparing} also conducted a quantitative evaluation of the usability of several crypto APIs. They conducted an online study with open source Python developers to test the following Python crypto libraries: cryptography.io, Keyczar, PyNaCl, M2Crypto, and PyCrypto. They tested symmetric and asymmetric encryption and key generation as well as certificate validation. Their  tasks were designed to be short enough so that the uncompensated participants would be likely to complete them before losing interest, but still complex enough to enable errors to be made. The authors conclude that, while cryptographic libraries should offer a simple and convenient interface, this is not enough. Developers also need to support a range of common tasks and provide accessible documentation with secure, easy-to-use code examples. However, due to the short task duration and online nature of the study, the authors could not do an analysis of the causes. The qualitative nature of our study, in combination with the much longer task, allows us to give a deeper analysis. In particular we show that task framing has a huge effect. 

Our study differs in several ways from the studies of Acar et al. In their first study they examined the effect of information sources on the security of the code. In their second study they analyzed the usability of crypto APIs for encryption. 
We are examining password storage and whether the task description and priming participants help in producing secure code. We are also looking at the difference between different APIs; however, we are using qualitative methods to gain deeper insights into the rationale behind developers' behavior. 

\section{Study design}
\label{sec:StudyDesign}
We designed a study through which we could gain insights into the development process of password storage and user authentication code. Specifically, we wanted to study the effect of telling participants that we want them to store the end-user passwords securely as opposed to them having to think of security on their own. Additionally, we wanted to see the effect of giving developers a framework which offers inbuilt functionality for securely storing passwords. Such a framework would remove the burden on the developer of having to choose a secure salt and hash function, as well as knowing that iterations are recommended. 

As is common practice in usability studies, we created a role-playing scenario to give the participants context. We asked the participants to imagine that they are working in a team to create a social networking platform and that they are responsible for creating the code for user registration and user authentication. We selected two different Web frameworks with different levels of support for secure password storage and divided the participants between these two frameworks (see Table \ref{table:useCases}). We refer to~\cite{finifter2011exploring}, where a taxonomy of categorization of levels that support a framework is explained in further detail. We chose frameworks with two levels of support in our analysis: (1) \textit{manual}, the weakest level of support in which developers have to write their own salting and hashing code using just crypto primitives, and (2) \textit{opt-in}, a framework which offers a secure implementation option, which can be used by developers if they think of it or find it. We did not find any framework which offers the level of \textit{opt-out} support for password storage.\footnote{This is probably due to the fact that it would be very hard to stop a developer from storing arbitrary strings in plain text. Our results show that this would, however, be a very desirable level of support to reach.}

In order to get substantive results we used multiple data-collection approaches for the study: (1) an implementation task which was logged in its entirety, (2) a survey, and (3) a semi-structured interview. Since we were interested in qualitative data, we opted to conduct the study  in a laboratory setting instead online. Although online studies have the advantage of letting people work in their natural environment, it is harder to monitor all their activity and the in-person interview is harder to conduct. In our usability lab, we provided the setting, including software and hardware, and thus were able to track all actions of the participants. Second, we had full control of time constraints and could avoid any kind of unnecessary disturbance during the implementation task. Finally, we could ensure that participants did not get outside help without such assistance being logged and the record available for analysis. 


Participants were briefed about the procedure and purpose of the study and told that they could quit at any time. Half the participants were told that the study was about the usability of Java frameworks, while the other half were told the study was about secure password storage. Thus, half the participants were subjected to a deception task. Only after completing the survey were these particular participants informed about the true purpose of the study. This deception was necessary since telling everyone that we were studying password storage would have primed those participants who had not been explicitly told to securely store the passwords, thus making it impossible to discover whether they had thought of secure storage on their own.  For those with the explicit task we did not need to use any deception. These participants were told the true purpose of the study in the \textbf{Introductory Text} and in the \textbf{Task Description}. Both groups completed a short pre-questionnaire about their expectation of how difficult the task would be and a self-assessment of their programming skills. After finishing the implementation task, participants completed a further survey and were interviewed about their implementation, their experience solving the task, and their thoughts on the usability of the framework they worked on. Finally, both groups were debriefed, with the non-primed participants additionally being informed about the true purpose of the study. 

In the following section, the rationales behind the choices made in the study design will be discussed. 
\begin{table*}[tb]
\centering
\begin{tabular}{ c c c c c }
    \toprule
    \textbf{No.} & \textbf{Framework} & \textbf{Level of Support} & \textbf{(Non-)Priming} & \textbf{Label}\\
    \midrule
    1 & JSF & manual & Priming & JP \\ 
    2 & JSF & manual & Non-Priming & JN \\
    3 & Spring & opt-in & Priming & SP \\
    4 & Spring & opt-in & Non-Priming & SN \\
    \bottomrule
\end{tabular}
\caption{Combination of 4 Different Scenarios in the Study}
\label{table:useCases}
\end{table*} 

\subsection{Language and Framework Selection}
Several statistics based on diverse analytical data indicate Java as one of the most widely employed server-side programming languages for applications and in the Web~\cite{GitHut, W3Techs, TIOBE, PYPL, RedMonk, TrendySkills}. 
Java is also the main programming language taught at our university; thus we chose Java as the programming language for our experiments.

To compare the different levels of support offered by frameworks and their APIs, we conducted an analysis of the most popular Web frameworks,
based on the data gathered from the open source platform HotFrameworks~\cite{HotFrameworks}.
HotFrameworks combines scores from GitHub~\cite{GitHub} (number of stars a git repository has for a framework) and Stack Overflow~\cite{StackOverflow} (questions tagged with a name of a framework) and creates a statistic of the most popular Web frameworks. Additionally, we used Google Trends~\cite{GoogleTrends} in order to verify the data offered by HotFrameworks. Google Trends provides a good overview regarding the development of interest in frameworks worldwide. We concentrated on frameworks that have been popular for the past five years.

We selected \textbf{Spring} and \textbf{Java Server Faces (JSF)} as the most suitable Web frameworks for our study in terms of popularity and level of support for secure end-user password storage as explained above. In order to store a password securely with Spring, the developer can make use of Spring Security's \texttt{PasswordEncoder} interface.
By contrast, participants implementing an application with JSF have to manually implement secure password storage or decide to integrate a library on their own. 


\subsection{Participants}
Researchers conducting computer science studies with professional developers often encounter obstacles, such as high costs and low response rate \cite{latoza2006maintaining}. In the domain of software engineering, there are multiple studies examining the applicability of students as proxies for developers~\cite{host2000using, berander2004using,Svahnberg:2008cj, salman2015students}. While there are caveats, on the whole it is considered acceptable to do exploratory research with students. Furthermore, the recent work of Acar et al. indicates that this holds true in the field of security software engineering~\cite{acar2016you, wermke2017security}. 



Therefore, we recruited 7 Bachelor and 13 Master students of computer science via the computer science mailing list and posters in the CS buildings (see Appendix \ref{sec:recruitment}). 
A demographics table can be found in the Appendix, Table \ref{demographics}. 

All applicants filled out a short questionnaire about their familiarity with Java and the APIs used in the study (see Appendix~\ref{sec:Pre-Screening}). 

Since the Bachelor program in Computer Science at the University of Bonn is completely in German and the Master program is in English, we gave our participants the opportunity to decide which language they wished to use for the study, including the written study task and the interview.
As a prerequisite for our study, familiarity with Java and the integrated development environment (IDE) Eclipse were required. 

\subsection{Task Design}
Our goal was to design a Web development task that was short enough to be solved in one working day but also long enough to hide the fact that secure password storage was the only thing of interest to us. We also wanted a longer task than the recent API comparisons of Acar et al. \cite{acar2017comparing} to observe participants having to potentially prioritize functionality or security. This is  similar to the real world, where functionality of an application is the primary task and security is the secondary task~\cite{whitten1999johnny, sasse2001transforming, sasse2003computer}. 


As previously mentioned, we asked participants to imagine they were part of a team working on creating a social networking site for our university. While our aim was not to quantitatively compare variables, we nonetheless studied four variants on how the study was conducted (see Table \ref{table:useCases}). We were interested in the qualitative feedback on Spring and JSF as well as when we directly asked students to implement secure password storage (henceforth called the primed group) and when they were not prompted (henceforth called the non-primed group).  


The application which the participants were supposed to extend was designed with the standard Model-View-Controller pattern. The front end was provided and the database back end fully implemented and documented. To ease the integration, students were given an illustration of the layers for the whole application and documentation of the existing Java classes with implementation hints (see Appendix \ref{sec:jsfhints}, \ref{sec:springhints}).
To facilitate a natural process, we explicitly allowed participants to use any kind of information source that could be found on the Internet.

Since there is still very little knowledge on how to design tasks for developer studies in the security realm, we conducted three pilot studies to refine the task design. The following three subsections briefly describe differences and improvements of the pilot studies.

\subsubsection*{\textbf{Pilot Study 1}}
In order to test our task design, we conducted a pilot study with 4 participants: 2 Computer Science PhD students of the Usable Security and Privacy group at the University of Bonn, and 2 Computer Science Master students in their 1st and 3rd semester at the University of Bonn. 
The main purpose of the first pilot study was to test timing, the understanding of the task, and to see if the students managed to complete the task. In this pilot study we only tested the priming version of the study since, a) the students knew our group and could not have been fooled into believing the non-priming scenario, and b) the three measurement goals can more reliably be achieved if students definitively attempt the security part of the task. 

We set the overall time threshold of the study to 8 hours. This time included a reception phase, an implementation task, a survey, an interview, a debriefing session, and a break of half an hour. Participants were compensated with 80 euros (10 euros per hour). We also asked our participants to ``Think-Aloud'', since this is a common method in end-user research to gain insights into the thoughts of participants. 

Participants had to write the code for the \textbf{registration} and an \textbf{authentication} part of the Web application with one of the two Web frameworks: Spring (PhD 1 and Master student 1) or JSF (PhD 2 and Master student 2). Registering an end-user means storing the user's data, which he/she provides through an interface to a database. Authenticating the end-user means to log in the user with his/her credentials provided in the registration process. 

Due to hardware problems, we had to exclude one participant (PhD student 2) from the study analysis. Participants indicated high skill levels in Java programming and Web development. On average, PhD student 1 and the first Master student needed about 7 hours to solve the task. The second Master student did not implement a functional solution for the \textbf{authentication} task within the 8 hours.

Based on our observations, we decided that our first design interaction was too complex and that students had difficulty completing the task in time. Because of this we decided to remove the authentication task and to focus entirely on password storage during the registration process.

We also found that, unlike in end-user studies, the Think-Aloud method did not provide us with valuable results. Most of participants did not talk at all during programming. If they were talking, they only addressed topics apart from our security focus. They stated obvious things, such as searching for a term, or items which were discussed in the interview after solving the task anyway. It seems that both the length and complexity of the task makes Think Aloud problematic. We therefore removed this element of the study. This also opened up the possibility of having multiple participants conduct the study at the same time.

All further studies were executed in our usability lab with 9 participants and 1 supervisor. The workspaces were set up in a way that participants could not see one another's work and they were equipped with noise canceling headphones to minimize distractions. 

\subsubsection*{\textbf{Pilot Study 2}}
In the second pilot study, participants had to implement the registration part of the Web application. As before, we only used the priming scenario. We invited 2 Computer Science Master students from the University of Bonn and 1 Computer Science Master student assistant from the Usable Security and Privacy group of the University of Bonn. While the student assistant was compensated with 60 euros (we expected that the implementation of the registration task would not exceed 6 hours), the Master students worked on the task in the course of a Usable Security and Privacy Lab. Thus, they were compensated with ECTS-credits\footnote{Participation was still voluntary and the student could have chosen a non-study task to comply with ethical guidelines.}. 

On average, participants took 5 hours and 30 minutes to complete the study. We observed that the database connection in particular was costing the participants a lot of time. Since this was not our primary interest, we added some hints for this part of the task so participants would not get lost in it. 

We estimated that the streamlined study would last about 4 hours. Since we wanted to recruit as many students as possible, we announced the next phase of the study as lasting approximately 4-5 hours, 8 hours max. (with a break of half an hour) and announced a payment of 60 euros. As the next subsection will show, this turned out to be a mistake.

\subsubsection*{\textbf{Pilot Study 3}}
We invited 7 participants to what was supposed to be the first round of the main study. They were randomly assigned to the priming/non-priming and Spring/JSF scenarios. On average participants took part in the study for about 5 hours and 30 minutes. However, we noticed that, unlike in the 8-hour pilot study, participants who did not complete the task within the estimated 4-5 hours got frustrated and gave up. We received non-functional and insecure solutions, and the interviews indicated that timing was a major factor for this. Also, unlike in the previous two groups, some participants did not know how to verify that they had solved the task. An interesting effect occurred in the non-primed group. After solving the task, participants were asked to complete a survey containing questions about secure password storage. Upon reading the questions, some participants went back into the code and implemented secure storage. 

Due to these critical issues, we discarded the results from these 7 students and created another iteration of the study design. Firstly, we revised the task description to make it clear when the task was functionally complete. 
The final version of the task description and implementation hints can be found in the Appendix \ref{sec:studyTask}, \ref{sec:jsfhints}, \ref{sec:springhints}. Secondly, we revised the time constraints on the notices for the study: \textit{``The study will last 8 hours.''} We decided to allow students to go back into the code after reading the questions in the survey but built in logging mechanisms to track when this happened. This actually gave us very valuable data that showed that the participants had the skills, but just did not think they were necessary. 

To give stronger incentives, we increased the payment for participation to 100 euros. Additionally, we provided snacks and refreshments to ensure participants' contentment. Previously, students had been asked to bring their own food. 

After this third pilot study the task design worked well, with the 20 students completing the study as planed. Our study was conducted in May 2017. 
The pilot studies were conducted in December 2016 and April 2017.

\subsection{Survey and Interview}
Once the implementation task had been finished or abandoned, participants had to fill out an exit survey. After that they were interviewed about their task solution and their knowledge about end-user password storage security mechanisms. The interview and survey questions can be found in the Appendix \ref{sec:survey} and \ref{sec:interview}.

\subsubsection*{\textbf{Survey}} For the survey we applied the suggested approach of~\cite{finstad2010response} to use seven-point rating scales instead of five-point scales. We asked participants for their demographics and programming experience. They also had to indicate their experience with the APIs employed in the study.

In order to address user satisfaction, we wanted to find out how easy or difficult the implementation task was perceived by participants (experience) in comparison to how easy or difficult they thought it was going to be (expectation), a method suggested in~\cite{albert2003you, tedesco2006comparison}. Therefore, we asked participants a \textit{Single Ease Question} (SEQ)~\cite{sauro2009comparison, SEQ} after they had read the study task description, but \textbf{before} they had started to work on the implementation task. Additionally, they had to answer the SEQ \textbf{after} they had finished the implementation task. 

\subsubsection*{\textbf{Interview}} After participants had completed the exit survey, they were interviewed about their implemented solutions and end-user password storage security mechanisms. They were also asked what they would do if they were working for a company and were given exactly the same task with the same description and time constraints. Specifically, they were asked whether they would have solved the task in the exact same way as they solved it in the study.

Regarding the priming and non-priming scenarios, both sets of interview questions sets were similar apart from those questions illustrated in Table \ref{table:priming-effect}. The primed participants, who were told to solve their task securely, were asked whether they would have done this if they had not been explicitly requested to do so. The non-primed participants were asked about security awareness in any case, regardless of how they solved the task. 

A single trained researcher conducted all 20 qualitative interviews. The interview sessions were audio-recorded. After the recordings were transcribed, two researchers independently coded all interviews using Ground-Theory and compared their final code books using the inter-coder agreement. The Cohen's ~\cite{cohen1960coefficient} kappa coefficient ($\kappa$)  for all themes was 0.81. A value above 0.75 is considered a good level of coding agreement ~\cite{fleiss2013statistical}.  

\begin{table}
\centering
\begin{adjustbox}{max width=\columnwidth}
\begin{tabular}{ P{1.25cm} P{3.65cm} P{3.65cm} }
\toprule
    & \multicolumn{2}{c}{\textbf{Participant states to have stored the end-user password}} \\
    & \textbf{\textit{securely}} & \textbf{\textit{not securely}} \\ \midrule
    \textbf{Primed Group} & ``Do you think you would have stored the end-user password securely, \textit{if you had not been told about it}''?  & - \\ \midrule
    \textbf{Non-Primed Group} & ``\textit{How did you become aware} of the necessity of security in the task? At which point did you decide to store end-user password securely?'' &  ``\textit{Were you aware} that the task needed a secure solution?'' \\
\bottomrule
\end{tabular}
\end{adjustbox}
\caption{Questions Asked Depending on the Priming /Non-Priming Scenarios}
\label{table:priming-effect}
\end{table}

\subsection{Evaluating Participants' Solutions}
For the usability analysis of the Spring and JSF APIs, we used the ISO usability specification as a basis, i.e., \textit{"the extent to which a product can be used by specified users to achieve specified goals with \textbf{effectiveness, efficiency and satisfaction} in a  specified context of use"}~\cite{standard1998ergonomic}. Both the \textit{survey and the interview} covered the satisfaction aspect. To specify the efficiency, we measured the \textit{time} a participant needed to solve the task. For specifying the effectiveness, we examined the \textit{functionality and security} of the participant's program code. If the end-user was able to register the Web application, meaning that his/her data provided through the interface was stored to a database, then the task was solved functionally to an adequate degree. In order to store end-user passwords securely, participants had to hash and salt the end-user passwords~\cite{grassi2017draft}.

Nowadays, GPUs can run billions  of  instructions  per second~\cite{orman2013twelve}. Consequently, the one-way hash functions \texttt{SHA1}~\cite{pub1994186, eastlake2001us} and \texttt{MD5}~\cite{rivest1992md5} can be calculated extremely quickly. Thus, these algorithms are no longer considered as sufficiently secure in most cases, as demonstrated by several serious password leaks within major organizations~\cite{finkle2012linkedin, kamp2012linkedin,  florencio2014administrator}. To counter this problem, more computationally intensive password hashing schemes (PHSs), like \texttt{PBKDF2 (Password Based Key Derivation Function)}~\cite{kaliski2000pkcs}, \texttt{bcrypt}~\cite{provos1999future} and \texttt{scrypt} \cite{percival2009stronger} have been proposed. These algorithms apply the idea of \emph{key stretching} to increase the computation time for testing each possible password (\emph{key}) by iterating the hash of the salted password~\cite{kelsey1997secure}. Due to the presence of cheap and parallel hardware, the widely used \texttt{PBKDF2} and \texttt{bcrypt} are vulnerable to massively-parallel attacks~\cite{orman2013twelve, hatzivasilis2015password, hatzivasilis2015lightweight}. In order to counter these attacks, memory-hard PHSs were introduced, since fast memory is considered expensive. \texttt{Scrypt} is a memory-hard PHS, but it is vulnerable to other types of attacks such as cache-timing and garbage-collector attacks~\cite{forler2013catena, forler2014memory}. By contrast \texttt{Argon2}~\cite{biryukov2015argon2}, the winner of the Password Hashing Competition (PHC)~\cite{PHC}, provides a defense against such attacks and can be viewed as the state of the art in the design of memory-hard functions. Other promising PHSs can be found in~\cite{hatzivasilis2015password, forler2014overview}.

Although we are aware of the fact that memory-hard functions like \texttt{scrypt} and \texttt{Argon2} are stronger than \texttt{bcrypt} and \texttt{PBKDF2}, we decided not to penalize participants for using current standards~\cite{grassi2017digital}. While NIST indicates the use of a suitable one-way key derivation function for password storage as a requirement to be followed strictly, the usage of memory-hard functions is recommended as particularly suitable. As an example for a suitable key derivation function \texttt{PBKDF2} is still suggested with an iteration count of at least $10\,000$. In previous specifications \cite{turan2010sp}, NIST even recommended an iteration count of at least $1\,000$. Since our study was conducted in May 2017, only a draft \cite{grassi2017draft} of NIST's current specification was available to our participants. Therefore, we accepted a solution with at least $1\,000$ iterations; however bonus points were awarded if participants changed the iteration count to at least $10\,000$.

In order to store a password securely with Spring, developers can make use of Spring Security's cryptography module \texttt{spring-
security-crypto}\footnote{since Spring Security 3.1}, which contains its own password encoder. Developers can simply import an implementation of the password encoder with a hash function they prefer. Although an implementation of \texttt{scrypt} is available, the official Spring Security Reference refers to \texttt{bcrypt} as the preferred hash function~\cite{SpringSecurityReference}. The provided implementation uses a cost parameter of 10 resulting in $2^{10}=1\,024$ key expansion iterations by default. Originally Provos and Mazieres proposed a cost parameter of 6 for end-user and 8 for administrator passwords~\cite{provos1999future}. While these values are considered outdated today, recent works use the cost parameter 12 for \texttt{bcrypt} as a common choice for analysis~\cite{hatzivasilis2015password, wiemer2014high, malvoni2014your, durmuth2014password}. By using Spring Security's \texttt{bcrypt} implementation, a random salt value of 16 bytes is integrated by default as well.
Again, we did not wish to penalize participants for using standards of frameworks, and so we rated parameter defaults of Spring Security's \texttt{bcrypt} implementation as secure.

By contrast, an application implemented solely with JSF can either be extended by Spring Security, or the developer can implement his or her own solution for secure password storage. For instance, he or she could make use of Java's Cryptography Extensions (JCEs). A random salt value has to be implemented by the developer as well. We did not restrict participants in their choice of how to implement secure password storage, neither for Spring nor for JSF.

To summarize, we rated security of end-user password storage under different assessment criteria. 
The highest possible score is 7 points, consisting as follows~\cite{grassi2017digital, PasswordStorageCheatSheet}:

\begin{enumerate}
    \item The end-user password is salted (+1) and hashed (+1).
    \item The derived length of the hash is at least 160 bits long (+1).
    \item The iteration count for key stretching is at least $1\,000$ (+0.5) or
    $10\,000$ (+1) for \texttt{PBKDF2} and at least $2^{10}=1\,024$ for \texttt{bcrypt}~(+1).
    \item A memory-hard hashing function is used (+1).
    \item The salt value is generated randomly (+1).
    \item The salt is at least 32 bits in length (+1).
\end{enumerate}

As explained above, in the following analysis we refer to solutions which achieved 6 \textit{or} 7 points as secure solutions. Solutions which earned 6 points followed industry best practices. In other words, from a usability perspective, these participants did all one could realistically expect from them. It must be stressed, though, that no participant received all 7 points and thus did not become aware of recent academic results showing that memory-hard hashing functions are necessary for best security ~\cite{biryukov2015argon2, hatzivasilis2015password}.

\begin{table*}[tb]
\begin{threeparttable}
\footnotesize
\centering
\begin{adjustbox}{max width=\textwidth}
\begin{tabular}{ c P{3cm}  P{3cm} P{3cm} P{3cm} P{3cm}}
\toprule
    \textbf{Participant} & \textbf{Security Expertise} & \textbf{Knowledge of Secure Password Storage} & \textbf{Stored Passwords Before} & \textbf{Framework supported me} & \textbf{Framework prevented me} \\ 
    
    & Not knowledgeable at all (4) - \newline Very knowledgeable (28)\tnote{1} & Not knowledgeable at all (1) - \newline Very knowledgeable (7) & & Strongly disagree (1) - Strongly agree (7) & Strongly disagree (1) - Strongly agree (7)
    \\ \midrule
    
    \rowcolor{LightGray}    JN1 & 15 & 3 & \xmark & 4 & 4 \\
    \rowcolor{LightGray}    JN2 & 12 & 2 & \checkmark & 4 & 4 \\
    \rowcolor{LightGray}    JN3 & 16 & 5 & \xmark & 3 & 3  \\
    \rowcolor{LightGray}    JN4 & 21 & 6 & \checkmark & 4 & 4  \\
    \rowcolor{LightGray}    JN5 & 13 & 3 & \checkmark & 4 & 4  \\
    JP1 &   13 & 2 & \checkmark & 3 & 1  \\
    JP2 &  23 & 5 & \checkmark & 2 & 4 \\  
    JP3 &  19 & 6 & \checkmark & 3 & 2  \\
    JP4 &  15 & 2 & \checkmark & 1 & 5  \\
    JP5 &   13 & 2 & \checkmark & 2 & 6 \\
    \rowcolor{LightGray}    SN1 &  16 & 6 & \checkmark & 1 & 1 \\
    \rowcolor{LightGray}    SN2 &  23 & 5 & \checkmark & 1 & 4 \\
    \rowcolor{LightGray}    SN3 &  17 & 4 & \checkmark & 4 & 4 \\
    \rowcolor{LightGray}    SN4 &  19 & 7 & \xmark & 1 & 7  \\
    \rowcolor{LightGray}    SN5 &   10 & 1 & \xmark & 4 & 4 \\
    SP1 &   15 & 4 & \xmark  & 7 & 1\\
    SP2 &   14 & 5 & \checkmark & 1 & 3 \\
    SP3 &    8 & 2 & \checkmark & 7 & 1  \\
    SP4 &    12 & 4 & \checkmark & 7 & 2 \\
    SP5 &   16 & 4 & \checkmark & 4 & 4 \\
\bottomrule
\end{tabular}
\end{adjustbox}
\caption{Participants' Security Expertise, Password Knowledge and Framework Support}
\begin{tablenotes}
\vspace*{-0.4cm}
\item[1]\footnotesize{We determined the security expertise from Q1-Q4 (see Appendix). We combined the answers of the participants to Q1-Q4 by adding up the numbers of their responses, whereby the answer for Q2 had to be reversed for consistency reasons. S=Spring, J=JSF, P=Priming, N=Non-Priming (See Table \ref{table:useCases} for labeling.) }
\end{tablenotes}
\label{table:securityExpertise}
\end{threeparttable}
\end{table*}

\subsection{Ethics}
Our institution does not have a formal IRB process for computer science studies, but the study protocols were cleared with the projects ethics officer. Our study also complied with the strict German privacy regulations.
The data was collected anonymously and our participants were provided with a consent form that had all information on their recorded data. The participants were informed about withdrawing their data during or after the study.

\section{Results}
We report statements of specific participants by labeling them JP/JN/SP/SN, from 1 to 5 each, according to their scenarios from Table \ref{table:useCases}. 
Table \ref{table:securityExpertise} summarizes participants' security expertise identified from the exit survey. The maximum security expertise to achieve is 28. Participants' security expertise ranged from 8 to 23. 

15 participants stored passwords before working on the study task. Despite this fact, many of them answered, in the question concerning their knowledge of secure password storage, that they have little understanding of this topic.

Further, participants identified their experience with the framework, which is discussed in section \ref{sec:Frameworks} in detail.

\subsection{Functionality and Security}
Table \ref{table:totalSolution} summarizes all solutions from our participants in terms of functionality, security, and implementation time. Not a single participant from the non-primed groups, either working with JSF or Spring, stored the end-user passwords securely in the database. The participants' implementation time in the non-primed groups varied from about 2 to 6.5 hours. From the primed groups, SP1, SP3, SP4, and JP4 were able to store the password securely with a security score of 6. However, none of the participants used a memory-hard hashing function and achieved all 7 points. SP1, SP3, and SP4 used \texttt{bcrypt} with a 192-bit password hash and a 128-bit salt. Their implementation time varied from about 3 hours to 7 hours. All of them made use of Spring Security's \texttt{PasswordEncoder} interface.  
 
JP2 and JP4 both used \texttt{PBKDF2} with different algorithms, iteration counts and salt lengths. Their implementation time varied from about 3 hours to 4 hours. JP2 employed \texttt{SHA256} with a hash length of 512 bits, $1\,000$ iterations and a 256-bit salt. JP4 used \texttt{SHA1} with a hash length of 160 bits, $20\,000$ iterations and a smaller salt of 64 bits. By contrast, JP3 used the hash function \texttt{SHA256} with only one iteration and without salting.

Some participants changed the length constraints of the password. JP1 and JP5 used 3-50 characters as length constraint for the end-user password. SN4 and SP1 decided on a password length greater than 6 characters, while SN1, SN2, and SP5 chose a password length constraint with a minimum of 8 characters. 

\begin{table*}[tb]
\footnotesize
\centering
\begin{adjustbox}{max width=\textwidth}
\begin{tabular}{ c P{1cm} P{2cm} | P{2cm}  P{1.5cm}  P{1.5cm}  P{1.5cm}  P{1.5cm} | P{1.5cm} }
\toprule
    & \multicolumn{1}{c}{\textbf{Time}} &\multicolumn{1}{c}{\textbf{Functionality}} & \multicolumn{5}{c}{\textbf{Security}} & \\
    & & & \multicolumn{3}{c}{\textbf{Hashing}} & \multicolumn{2}{c}{\textbf{Salt}} & \\
    & (hh:mm) & Storage working & Hashfunction \newline (at most +2) & Length (bits) \newline(+1 if $\geq$ 160) & Iteration count \newline(at most +1) & Used library \newline (at most +2) & Length (bits) \newline(+1 if $\geq$ 32) & \textbf{Total} \newline (7)
    \\ \midrule
    \rowcolor{LightGray}    JN1 & 06:26 & \xmark &  &  &  &  &  &   0 \\
    \rowcolor{LightGray}    JN2 & 04:05 & \checkmark &  &  &  &  &  &   0 \\
    \rowcolor{LightGray}    JN3 & 03:01 & \checkmark &  &  &  &  &  &   0 \\
    \rowcolor{LightGray}    JN4 & 04:11 & \xmark &  &  &  &  &  &   0 \\
    \rowcolor{LightGray}    JN5 & 05:30 & \xmark &  &  &  &  &  &   0 \\
    JP1 & 04:55 & \checkmark &  &  &  &  &  & 0 \\
    JP2 & 03:12 & \checkmark & PBKDF2(SHA256) & 512 & $1\,000$ & SecureRandom & 256 & 5.5 \\
    JP3 & 05:29 & \checkmark & SHA256 & 256 & 1 &  &  &   2 \\
    JP4 & 04:12 & \checkmark & PBKDF2(SHA1) & 160 & $20\,000$ & SecureRandom & 64  & 6 \\
    JP5 & 06:32 & \checkmark &  &  &  &  &  & 0 \\
    \rowcolor{LightGray}    SN1 & 03:15 & \checkmark &  &  &  &  &  & 0 \\
    \rowcolor{LightGray}    SN2 & 02:24 & \checkmark &  &  &  &  &  & 0 \\
    \rowcolor{LightGray}    SN3 & 02:01 & \checkmark &  &  &  &  &  & 0 \\
    \rowcolor{LightGray}    SN4 & 04:01 & \checkmark &  &  &  &  &  & 0 \\
    \rowcolor{LightGray}    SN5 & 04:50 & \checkmark &  &  &  &  &  & 0 \\
    SP1 & 03:15 & \checkmark & BCrypt & 192 & $1\,024$ & Spring Library & 128  & 6 \\
    SP2 & 01:54 & \checkmark & MD5 & 128 & 1 &  &  &  1 \\
    SP3 & 07:00 & \xmark & BCrypt & 192 & $1\,024$ & Spring Library & 128 &  6 \\
    SP4 & 03:39 & \checkmark & BCrypt & 192 & $1\,024$ & Spring Library & 128 &  6 \\
    SP5 & 03:44 & \xmark &  &  &  &  &  & 0 \\
    
\bottomrule
\end{tabular}
\end{adjustbox}
\caption{Evaluation of the Implemented Solutions}
\label{table:totalSolution}
\begin{tablenotes}
\centering
\vspace*{-0.4cm}
\item[1]\footnotesize{S=Spring, J=JSF, P=Priming, N=Non-Priming (See Table \ref{table:useCases} for labeling.)}
\end{tablenotes}
\end{table*}

\subsection{Choice of Hash Function}
Our participants chose hash functions based on different factors. SP1 noted that he chose to work with \texttt{bcrypt}:

\begin{quote}
``A lot of forums say that is very secure and very up to date and
has a high vote in some cryptography website. [...] I find it trust-able, the Stack Overflow and cryptography forums.''
\end{quote}
In contrast, JP2 decided to not use \texttt{bcrypt} as he did not trust the library source. 
Further reasons for choosing a library were past experience. SP2 decided on \texttt{MD5} because he had used it before. JP4 would have also used \texttt{MD5} if the task had not asked for security:
\begin{quote}
``I would have used \texttt{MD5} because I was already comfortable with it and so I wouldn't have thought it had vulnerability.''
\end{quote}

\subsection{Expectations and Experiences With the Task} 
In order to evaluate the task difficulty before and after implementation, we asked our participants to rate their expectations and experiences.
The table summarizing the before and after ratings can be found in the Appendix (see Table \ref{table:expectations}). We additionally report the mean values of expectation and experience with regard to the priming/non-priming and framework support scenarios. However, these values have to be regarded with care since the main purpose of qualitative research is to explore a phenomenon in depth and not to generate quantitative results\footnote{It should be considered that our sample is small and not all groups indicate a normal distribution. The individual values can be found in Table \ref{table:expectations} (see Appendix).}. 
On a seven-point rating scale (1: Very difficult - 7: Very easy) the mean values for the non-primed group were 4 before and 4.1 after working on the task. The mean values for the primed group were 3.7 before and 3.8 after working on the task.

Before the implementation phase, our primed participants rated the task difficulty slightly more difficult than the non-primed participants. This might indicate that our participants acknowledged the security overhead to increase the difficulty of the task. This bears further examination in the future since such a perception might make developers shy away from security code. 

Comparing the mean values of the framework support scenarios, the groups had nearly similar mean values (Spring: 3.8, JSF: 3.9) before the task implementation. After the task, however, the mean value of experience changed to 4.3 for Spring and 3.6 for JSF. This might indicate that the participants perceived solving the task with JSF as more complex than solving it with Spring.

Further, we observed that participants who implemented solutions with a low security score (JP3, SP2), e.g. because of using weak hash functions, rated the difficulty of the task as rather low. SP2 stated that he used \texttt{MD5} because he had already used it in previous applications. His self-reported familiarity with the task could explain his rating. 
All other participants who were able to securely implement the task indicated it as more difficult. 

We observed no noticeable trends in the expected difficulty and performance of the participants. The participants from the primed group who added (limited) security gave a similar rating (JP3, SP1, SP3) or even a slightly better rating (JP2, JP4, SP2, SP4) after implementation. However, the participants who were not able to store passwords securely rated the task slightly more difficult (JP1, JP5) after the implementation phase. This indicates that they underestimated the complexity of the security part of the task.

\begin{table*}[tb]
\footnotesize
\centering
\begin{adjustbox}{max width=\textwidth}
\begin{tabular}{ c P{1.875cm} P{1.875cm} | P{1.875cm}  P{1.875cm}  P{1.875cm}  P{1.875cm}  P{1.875cm} P{1.875cm} }
\toprule
    & \multicolumn{2}{c}{\textbf{Secure storage}} & \multicolumn{2}{c}{\textbf{Hashing}} & \multicolumn{2}{c}{\textbf{Salting}}  & \multicolumn{2}{c}{\textbf{Company}} \\
    & Security-Score \newline (7) & Own Opinion & Basic Understanding & Deeper Understanding & Deeper Understanding & Believed Have Used & Similar Solution & Further Code Improvement \\ \midrule
    \rowcolor{LightGray}    JN1 & 0   & \xmark     & \xmark     & \xmark     & \xmark          & \xmark              & \checkmark & \checkmark \\
    \rowcolor{LightGray}    JN2 & 0   & \xmark     & \checkmark & \xmark     & \xmark          & \xmark              & \xmark     & \checkmark \\
    \rowcolor{LightGray}    JN3 & 0   & \xmark     & \checkmark & \xmark     & \checkmark      & \xmark              & \xmark     & -\\
    \rowcolor{LightGray}    JN4 & 0   & \xmark     & \checkmark & \checkmark & \checkmark      & \xmark              & \xmark     & \checkmark \\
    \rowcolor{LightGray}    JN5 & 0   & \xmark     & \checkmark & \xmark     & \xmark          & \xmark              & \xmark     & \checkmark \\
                             JP1 & 0   & \xmark     & \checkmark & \xmark     & \xmark          & \xmark              & -          & - \\
                             JP2 & 5.5   & \checkmark & \checkmark & \checkmark & \checkmark      & \checkmark          & \xmark     & \checkmark \\
                             JP3 & 2   & \checkmark & \checkmark & \checkmark & \checkmark      & \checkmark          & \checkmark & \checkmark \\
                             JP4 & 6   & \checkmark & \checkmark & \checkmark & \textbf{\xmark} & \textbf{\checkmark} & \checkmark & - \\
                             JP5 & 0   & \xmark     & \checkmark & \checkmark & \xmark          & \xmark              & \checkmark & - \\
    \rowcolor{LightGray}    SN1 & 0   & \xmark     & \checkmark & \checkmark & \checkmark      & \xmark              & \checkmark & - \\
    \rowcolor{LightGray}    SN2 & 0   & \xmark     & \checkmark & \checkmark & \checkmark      & \xmark              & \checkmark & \checkmark \\ 
    \rowcolor{LightGray}    SN3 & 0   & \xmark     & \checkmark & \checkmark & \checkmark      & \xmark              & \checkmark & \checkmark \\ 
    \rowcolor{LightGray}    SN4 & 0   & \xmark     & \checkmark & \checkmark & \checkmark      & \checkmark          & \xmark     & \checkmark \\ 
    \rowcolor{LightGray}    SN5 & 0   & \xmark     & \checkmark & \checkmark & \xmark          & \xmark              & \checkmark & \checkmark \\ 
                             SP1 & 6   & \checkmark & \checkmark & \checkmark & \checkmark      & \checkmark          & \checkmark & - \\ 
                             SP2 & 1 & \checkmark & \checkmark & \checkmark & \checkmark      & \xmark              & \xmark     & \checkmark \\ 
                             SP3 & 6   & \checkmark & \xmark     & \xmark     & \xmark          & \checkmark          & \checkmark & \checkmark \\ 
                             SP4 & 6   & \checkmark & \checkmark & \checkmark & \checkmark      & \checkmark          & \checkmark & \checkmark \\ 
                             SP5 & 0   & \xmark     & \checkmark & \xmark     & \xmark          & \xmark              & \checkmark & \checkmark \\ 
    
\bottomrule
\end{tabular}
\end{adjustbox}
\caption{Participants' Security Knowledge and Code Self-Assessment }
\label{table:securityKnowledge}
\begin{tablenotes}
\centering
\vspace*{-0.4cm}
\item[1]\footnotesize{S=Spring, J=JSF, P=Priming, N=Non-Priming (See Table \ref{table:useCases} for labeling.)}
\end{tablenotes}
\end{table*}

\subsection{Code Self-Assessment vs. Actual Security}
The participants' code self-assessment sometimes differed from the solutions they actually implemented. 
Table \ref{table:securityKnowledge} summarizes the security knowledge of each participant and the participants' self-assessment regarding their own solution we learned from the interviews compared to our security-score. 
Furthermore, 
their estimations on whether they would solve the task differently in a real company are listed in the table.

Lacking knowledge of hashing and salting does not necessarily lead to insecure code, nor vice versa. With the exception of SP3 and JN1, all participants presented at least a basic knowledge of hashing. A number of participants (JN1, JN2, JN5, JP1, JP4, JP5, SN5, SP3, SP5) did not know the definition of salting. The other participants explained that salting prevents usage of rainbow tables, thereby adding an additional level of security. While JP3 and SP2 assumed that they implemented the password storage securely, they in fact did not. SP3 was not able to explain hashing or salting correctly but managed to store the passwords securely by using the Spring helper classes. He stated:
\begin{quote}
``Every time you put a password in it, e.g. `password', you get each time another hash.''
\end{quote}
Although his implementation was not functional and his understanding of hashing not correct, he was able solve the task securely.
By contrast, the participants from the non-primed Spring group all showed a deeper understanding of hashing during the interviews, but did not store the end-user password securely. Also, in the primed Spring group, SP2 for instance could explain hashing and salting properly, but only gained a security score of 1/7.


What is more, in our pilot studies some participants mentioned that they would have implemented a different solution if the task was assigned in an actual company. This is a problem many studies face. To get insights into this we asked our participants whether their solution from the study would differ from the one they would implement for a company. 

Almost all participants stated that they would have solved the task in a better way. SP3 and JN5 stated that in a company they would ask for help.
Interestingly, some participants explained that there would be a security expert or supervisor who would check for and require security: 
\begin{quote}
``I would ask my supervisor about it. [...] There is definitely another person that understood these kinds of things.'' (JN3)
\end{quote}
The participants expected somebody to set security as a requirement or explicitly request it.
SN3 and JN4 mentioned that the security of their solutions depend on the company:

\begin{quote}
``It depends on the company. If it had been a security company I would have thought of something because they would have minded.'' (SN3)
\end{quote}
JN4 goes even further saying that it depends on the data the company is handling:
\begin{quote}
``Depends on the company data. If the data would be e.g. medical data, there would be other requirements.'' (JN4)
\end{quote}
SP2 also thinks the responsibility lies with the employer or the company itself. He stated:
\begin{quote}
``If the employer did not really constrain the programmer with certain requirements, then the programmer will do the work in, let's say, not in a sufficient way.''
\end{quote}
Other participants stated that if they had more time to get used to working with the framework, they would be able to produce a better solution.

\subsection{Experienced Spring/JSF Support}
\label{sec:Frameworks}
Table \ref{table:securityExpertise} shows the agreement of all participants on whether the framework they used supported or prevented them from storing passwords securely.

7 of the non-primed participants specified an undecided rating in the middle of the scale. SN3, for example, augmented his answer with these words: 
\begin{quote}
    ``I would say, that I'm undecided or don't know if it would have supported me, because I didn't look for it''
\end{quote} And to the prevention-related question he stated: ``Actually the same answer.''

Particularly interesting are the answers of participants belonging to the primed group regarding the framework support. While 3 of the 5 participants who were using Spring, managed to store the passwords securely, and thus, agreed that Spring supported them (SP1, SP3, SP4), all participants from the primed group using JSF disagreed that JSF supported them. Most participants that used JSF even answered that the framework prevented them from storing the password securely. JP4 mentioned:
\begin{quote}

    ``I thought of finding something within the JSF Framework or within a Java framework which easily gives a hash function or they can use it easily but I couldn't get anything.''
\end{quote}

SP1, for example, in answering why the Spring Framework supported him, said: 

\begin{quote}
    ``Because it has the built in library \texttt{bcrypt} for example and some strong validation framework for the password.''
\end{quote}

This observation could suggest that developers, who are told that a secure storage is a requirement, first look for a library in the framework they are working with and - if they find it - are willing to use it for the task.

Other participants gave Spring as the reason for why they did not manage to store the end-user password securely. SN4 explained:
\begin{quote}
``Because there are too many possibilities to implement user password hashing and Spring works in a way that it uses an authentication manager or something specific to Spring and one needs to extend the application to create this encryption or hashing schema.''
\end{quote}
SN4 even had a bad experience with Spring in his past job and stated that it was the main reason he quit the job. Further, SN2 felt that Spring hindered him from storing passwords securely because he did not know where to include his code into the framework. Another participant noted that JSF supported secure password storage ``Because it provides that validation thing'' (JN5). 

Some participants were not familiar with Hibernate and experienced difficulties with inserting data from the form to the database (JP2, JN5).  
JP3 had problems with MVC in general, and had to work into it before working with it. 
Further, some participants needed more time to get familiar with the frameworks (SN4, SN1).

\subsection{Priming vs. Non-Priming Scenario}
None of the 10 non-primed participants implemented a secure solution. The reasons for this outcome are many and varied. 
Most of the participants mentioned that the task description did not require it, or rather they did not know it was a requirement (JN2, JN3, JN5, SN1, SN2, SN3, SN4, SN5). While this may seem obvious, there is still a potentially important lesson to learn from this. Even though we would expect that developers should realize that storing passwords is a security critical task, some did not even make that connection, as the following quotes indicate: 

\begin{quote}
``Umm, actually literally when I was in the project I didn't feel much like that it was related to security.''~(JN5)
\end{quote}

\begin{quote}
``I thought the purpose of the study is not about security so I didn't do it.''~(JN2)
\end{quote}
Even if JN5 was aware of security in general he did not attempt to store the end-user password securely, because his task did not explicitly ask him to do so. Others did not think of security at all. For instance, JN2 noted: 
\begin{quote}
``It was like just to store information in tables and I didn't notice that, I should store password securely.''
\end{quote}
During the completion of the exit survey, some participants recognized that secure password storage might be part of their task. Participants were asked, whether they believe they had stored the end-user password securely. JN2, e.g., noted:
\begin{quote}
``When I completed the task until then I didn't notice this, but when I started filling the form, the survey form at the end of the task, then I felt that I missed this point.''
\end{quote}

What is more, one non-primed participant struggled with the Spring Framework. He reasoned it was hard to find a way to solve the task securely ``because there are too many possibilities to implement user password hashing...'' and that he ``would need more time to get familiar with the system'' (SN4).

By contrast, 7 of 10 primed participants at least hashed the passwords and stored them on the database. SP1 explained that even if he would not have been primed he would have stored the passwords securely, since his colleague experienced the downsides of insecure password storage.
SP2 also claimed that he would have stored the end-user passwords securely even if he had not been told to do so. However, he only used \texttt{MD5} to secure the passwords.
Finally, when JP1 was asked why he stored the password in plain text, he gave this candid answer:
\begin{quote}
 ``I'm lazy.''
\end{quote} 

This represents one of the main findings of our work. Even in a task so obviously security-related, participants who were not explicitly told to securely store the end-user passwords did not. While part of this effect can be explained by the fact that the participants knew that they were in a study, it is still important to note that not all of our participants linked password storage to a security-sensitive task. Thus we find that, similar to misperceptions end-users have about their password behavior, developers have misperceptions of their own, which we need to combat. Section \ref{miscon} goes into more detail on this matter.

Another important take-away is that developer studies need to take care when framing their tasks, since we saw a very large effect. Giving developers small focused tasks concerning crypto libraries will not reveal all the problems we saw in the more realistic context of complex tasks which mix security and functionality development. Both types of studies are worthwhile; however, the decision needs to be taken consciously according to the goal of the studies. 

\subsection{Misconceptions of Secure Password Storage}\label{miscon}
When we asked our participants what storing passwords securely means, some stated that the passwords need to have certain constraints and be validated for security, e.g. character length (JN5).
It became clear that some participants assume that securing passwords includes validation of the passwords on the end-user side instead of secure storage in the database on the developer side. 
What is particularly interesting is that developers were applying end-user security advice but not developer security advice. 
Another interesting finding is that one participant said that her solution is optimal because she tried to switch on SSL and secure the password transmission. JN1 saw no reason for hashing the passwords, stating:
\begin{quote}
 ``I assumed that the connection will be a secure connection like with an HTTPS connection, so everything should come encrypted.''
\end{quote}
Further, some participants had never heard of, let alone could explain, hashing or salting a password (JN1, SP3).

SP2 was sure that \texttt{MD5} is secure. He was not aware of the consequences of using the broken hash function and even stated:
\begin{quote}
``\texttt{MD5} with complicated salt is pretty secure.''
\end{quote}

\section{Limitations}
Below, we discuss the limitations of our study.\\
\textbf{Population:} Firstly, we used only computer science students instead of real developers. While related work from the field of empirical software engineering shows that students can be good proxies for real developers, it is still important to remember that there are differences between the two groups. Particularly, we want to underline that a lot of software development skills come from experience, which can be markedly absent from students.
Secondly, all students were recruited from the same CS environment. While such a homogenous group enables a better comparison between the different conditions, it limits generalizability. Thus, no generalizability is claimed. 
Thirdly, despite 20 participants being a good number for a qualitative study, the high variability of skill and experience levels means that our results need to be backed up by follow-up work with larger scale studies. 

\textbf{Laboratory environment:} The laboratory environment which allowed us to gather very rich and in-depth data also introduced an environmental bias. Participants knew they were part of a study and not working with real data. We used the common technique of role-playing to counter this as best we could; however, some participants nonetheless self-reported that they would have behaved more securely if they were given the same task in a company. We cannot quantify the effective size of this bias,  since it is also possible that such claims were made out of embarrassment to cover for mistakes. 


\textbf{Programming language:} We selected Java as a programming language to ensure that our entire population sample had expertise in it. It is the main teaching language at our university, so our students are all familiar with it. It is entirely possible that other programming languages and libraries would produce different results. Further work is needed to make any kind of generalizable statements. 

\textbf{Study length:} We would like to point out the length of the study as another limitation. We picked a whole working day as the length due to the complexity of the task, but refrained from using a multi-day time frame, since it would introduce uncertainty as to what participants did in their off-hours. Acar et al. show that the used information sources have a big effect and we need to be able to capture that~\cite{acar2016you}. We conducted extensive pilot studies to tailor the tasks to the time available and did not note any fatigue effects. Nonetheless, the study length is sure to have an effect - in particular in comparison to the very short studies carried out in related work~\cite{acar2016you, wermke2017security, acar2017comparing, stylos2008implications}. Whether short, one-day, or multi-day studies are more ecologically valid is unknown and must be examined further in future work. 

Overall, at present this study should be seen as the first step to offer insights into an extremely complex problem, and the results need to be interpreted as such. Our results need to be replicated and further methodological research is called for. All the above limitations need to be studied and expanded on. In particular, future work should examine different types of developers such as volunteers from online forums, for-hire developers, local developers, etc. and different task descriptions and task settings. 

\section{Discussion}
Developers are responsible for the software they implement. Interestingly, a number of our participants did not feel responsible for security when it was not explicitly mentioned. This is a surprising similarity to end-users who see security as a secondary task. 
Getting developers to care about security is essential in order to produce software which protects its users. However, our study shows that there is still much we do not know on why developers behave the way they do. In particular, the insights gained from the qualitative approach proved to be enlightening to us since it goes beyond simply quantifying where the problems lie. 

Some participants used weak hash functions like \texttt{MD5} or \texttt{SHA256} without key stretching simply because they were either not aware of the issues or had no negative experiences with them. 


People tend to use the hash functions, they are familiar with - even if they have heard about more secure hash functions. 
Although in general non-primed participants did not consider secure password storage, most of them showed a deeper understanding of it when prompted in the interviews. They were able to explain in detail how they would hash and salt an end-user password. Additionally, most participants were able to identify hash algorithms that were less recommended by referring to community discussions.

None of our participants used a memory-hard hashing function to store the end-user passwords securely, even though Spring Security offers an implementation of \texttt{scrypt} and participants could have used it just as easily as \texttt{bcrypt}.
This highlights the need for academics to intensify the push of research results into tutorials, standardization bodies, and frameworks. Another solution worth exploring is removing less secure options from libraries. This solution needs to be studied with care since it creates issues of backward compatibility which might create worse security problems in the long run. However, the fact that not a single participant used state of the art memory-hard hashing functions is a troubling discovery and one that calls for urgent action. 


Secure password storage was regarded from different perspectives in our study. Some participants thought their job was to ensure that end-users behaved securely, e.g. by setting constraints to password length and strength. Other participants thought that using encrypted transmission would achieve secure password storage. 
This highlights that developers have similar security misconceptions as end-users. Even though developers show greater theoretical knowledge of security concepts, this is no guarantee that they will use them or even acknowledge their necessity in their software.

Overall, our study indicates that, even when frameworks with better security support can help, priming is essential to produce secure applications. Developers need to be told explicitly about what kind of security to add and they need help in avoiding outdated security mechanisms.

\section{Conclusion}
It is important to find out why some developers store end-user passwords securely, while others do not. One assumption is that the amount of support the API offers to developers has an impact on secure password storage. However, our study suggests that API usability is not the main factor; in addition to the factor of documentation discovered by Acar et al.~\cite{acar2016you}, task framing and developer misconceptions play a very important role in the (in)security of code created by developers. 

We designed a laboratory study with four scenarios in order to analyze the different behaviors and motives of developers when implementing a password storage task. We selected two popular Web application frameworks with different levels of built-in support for secure password storage, Spring and JSF. 
Additionally, one group was told that the purpose of the study was to examine the usability of Java frameworks while the other group was explicitly told the study was about security and was directed to store the passwords securely. The key observations from our study are as follows: 

\textbf{Developers think of functionality before security.}
We conducted semi-structured interviews immediately after the participants finished working on the implementation task. The interviews provided us valuable insights into developers' perceptions of task accomplishment. Just as security is a secondary task for most end-users, our developers stated that they always concentrated  on the functionality of the task first before (if at all) thinking of security. 

\textbf{Asking for security can make a difference.} Our results indicate that there are big differences if a developer is explicitly asked to consider security. None of our non-primed participants thought that the task of storing passwords needed a secure solution, despite password storage obviously being a security sensitive task.  By contrast, participants who were advised to store the end-user password securely used various hash functions with different configurations. The configurations varied in their level of security. Most participants of the primed Spring group made use of Spring Security's \texttt{PasswordEncoder} interface. Thus, they were less likely to make mistakes.

\textbf{Standards and advice matter.} None of the participants who attempted to implement secure password storage produced a solution that met current academic standards. This needs to be seen as a call to action for academics to find better ways to disseminate the latest results in standards and tutorials as well as integrate them with popular frameworks. For the latter option, the trade-off between backward compatibility, flexibility, and security needs to be re-examined.  

\textbf{Opt-out security.}
At present, even frameworks which have high-level support for secure password storage offer this in an opt-in manner, i.e. developers have to know and want to use this feature. Based on the many misconceptions and lack of primary-task motivation, it would be beneficial to offer safe defaults. This is an open research problem, since it is non-trivial to recognize when developers are writing passwords in plain text as opposed to other strings across frameworks and IDEs. Ensuring that developers cannot accidentally store a password in plain text or with weak security is a difficult challenge, but one worth tackling. 

This study only presents a first qualitative look at password storage as a research problem. In future work, studies are needed to validate and extend the results. Firstly, our student sample limits the generalizability; thus, studies involving programmers with different skills and backgrounds from companies as well as freelancers are needed. Secondly, it is necessary to increase the sample size, in order to gather quantitative data.
Thirdly, once the problem is fully understood, work needs to begin on creating more usable password storage solutions to help developers with this critical task. 

\section*{Acknowledgements}
This work was partially funded by the ERC Grant 678341: Frontiers of Usable Security.

\bibliographystyle{ACM-Reference-Format}
\bibliography{references}

\section*{Appendix}

\begin{appendix}
\footnotesize
\section{Recruitment Invitation}
\label{sec:recruitment}
\textbf{Participant invitation for scientific studies}\\
Dear computer science students, We are conducting several scientific studies, in which you can participate and earn
100 euros!\\
\textbf{Details:}\\
The institute for Computer Science of the University of Bonn is looking for motivated Computer
Science Students (over 18), who want to take part in a scientific study with the topic of Web
Development in Java and to earn some extra money. The goal of the study is to test the usability of
different Java web development APIs. In the study you will be asked to implement parts of a web
application. Basic knowledge of Java and the IDE Eclipse is required. The study will last 8 hours.
Afterwards we will conduct a short interview and ask some questions about the tasks. The interview
will be audio recorded to facilitate the evaluation of results. The aim of the study is not to test your
knowledge but the usability of the APIs. All data gathered during the study will be anonymized.
Anonymized data and quotes may be published as part of a scientific publication. Your consent will be requested on the day you participate.
You will be payed 100 euros for taking part in the study.
Knowledge required: IDE Eclipse, JAVA
Interested?
Please fill in the following questionnaire to register your interest: \textit{LINK}
We are looking for a good mix of skills, so please fill out the questionnaire as accurately as
possible. We are looking for up to 120 participants. Registering your interest does not guarantee
participation.

\section{Pre-Screening Questionnaire}
\label{sec:Pre-Screening}
\begin{itemize}
\item Gender:
    \begin{itemize}
    \item Female, Male, Other, Prefer not to say
    \end{itemize}
\item Which university are you at?
    \begin{itemize}
    \item University of Bonn
    \item Other: 
    \end{itemize}
\item In which program are you currently enrolled?
\begin{itemize}
    \item Bachelor Computer Science
    \item Master Computer Science
    \item Other:
\end{itemize}
\item Your semester:
\item How familiar are you with Java?\\ \textbf{1: Not familiar at all - 7: Very familiar}
\item How familiar are you with PostgreSQL?\\  \textbf{1: Not familiar at all - 7: Very familiar}
\item How familiar are you with Hibernate?\\  \textbf{1: Not familiar at all - 7: Very familiar}
\item How familiar are you with Eclipse IDE?\\  \textbf{1: Not familiar at all - 7: Very familiar}
\end{itemize}
\small
\begin{table*}[tb]
\centering \begin{adjustbox}{max width=\textwidth, max totalheight=\textheight} \begin{tabular}{ccccccccccccc}
\toprule
 & & & & & & & \multicolumn{4}{c}{\textbf{How familiar are you with}}  \\
 & & & & & & & \multicolumn{4}{c}{Not familiar at all (1) - Very familiar (7)}  \\

&\textbf{Gender}&\textbf{University}&\textbf{Study program}&\textbf{Age}&\textbf{Nationality}& \textbf{Semester} &\textbf{Java} &\textbf{PostgreSQL}&\textbf{Hibernate} &\textbf{Eclipse IDE} &\textbf{Total skills} \\
\midrule

\rowcolor{LightGray} JN1&Male&University of Bonn&MSc Computer Science&NA&Bangladeshi&8&6&4&4&6&20&\\ 
\rowcolor{LightGray} JN2&Female&University of Bonn&MSc Computer Science&23&Pakistani&3&3&1&1&6&11&\\ 
\rowcolor{LightGray} JN3&Male&University of Bonn&MSc Computer Science&25&Uzbek&2&5&3&2&5&15&\\ 
\rowcolor{LightGray} JN4&Male&University of Bonn&BSc Computer Science&23&German&6&5&2&1&6&14&\\ 
\rowcolor{LightGray} JN5&Female&University of Bonn&MSc Computer Science&27&Indian&5&5&4&1&6&16&\\ 
JP1&Male&University of Bonn&MSc Computer Science&25&Chinese&5&5&1&1&4&11&\\ 
JP2&Male&University of Bonn&BSc Computer Science&22&German&4&6&6&1&6&19&\\ 
JP3&Male&University of Bonn&MSc Computer Science&26&Iranian&4&4&2&2&6&14&\\ 
JP4&Male&Aachen University&MSc Media Informatics&27&Indian&2&4&2&1&3&10&\\ 
JP5&Male&Aachen University&MSc Media Informatics&25&NA&2&2&1&1&2&6&\\ 
\rowcolor{LightGray} SN1&Male&University of Bonn&MSc Computer Science&24&German&10&6&4&1&5&16&\\ 
\rowcolor{LightGray} SN2&Male&University of Bonn&BSc Computer Science&20&German&2&7&5&2&4&18&\\ 
\rowcolor{LightGray} SN3&Male&University of Bonn&BSc Computer Science&24&German&8&6&3&1&6&16&\\ 
\rowcolor{LightGray} SN4&Male&University of Bonn&MSc Computer Science&25&Syrian&3&7&5&7&7&26&\\ 
\rowcolor{LightGray} SN5&Male&University of Bonn&BSc Computer Science&19&German&2&5&4&1&4&14&\\ 
SP1&Male&University of Bonn&MSc Computer Science&25&NA&4&4&3&2&4&13&\\ 
SP2&Male&University of Bonn&MSc Computer Science&25&Syrian&4&6&3&4&4&17&\\ 
SP3&Male&University of Bonn&BSc Computer Science&20&German&2&5&3&1&4&13&\\ 
SP4&Male&University of Bonn&BSc Computer Science&25&German&10&5&3&1&5&14&\\ 
SP5&Female&University of Bonn&MSc Computer Science&NA&Indian&4&5&4&4&6&19&\\ 
\bottomrule
\end{tabular} \end{adjustbox} \caption{Participants' Demographics} \label{demographics}
\begin{tablenotes}
\centering
\vspace*{-0.4cm}
\item[1]\footnotesize{S=Spring, J=JSF, P=Priming, N=Non-Priming (See Table \ref{table:useCases} for labeling.)}
\end{tablenotes}
\end{table*} 
\begin{table}[tb]
\footnotesize
\centering
\begin{adjustbox}{max width=\textwidth}
\begin{tabular}{ c c c }
\toprule
    & \textbf{Expectation} & \textbf{Experience}\\ 
    & \multicolumn{2}{c}{Very difficult (1) - Very easy (7)}\\ \midrule
\rowcolor{LightGray}    JN1&5&5\\ 
\rowcolor{LightGray}    JN2&4&4\\ 
\rowcolor{LightGray}    JN3&3&4\\ 
\rowcolor{LightGray}    JN4&5&3\\ 
\rowcolor{LightGray}    JN5&3&2\\ 
JP1&3&2\\ 
JP2&4&5\\ 
JP3&5&5\\ 
JP4&2&4\\ 
JP5&4&2\\ 
\rowcolor{LightGray}    SN1&4&6\\ 
\rowcolor{LightGray}    SN2&5&4\\ 
\rowcolor{LightGray}    SN3&3&5\\ 
\rowcolor{LightGray}    SN4&5&5\\ 
\rowcolor{LightGray}    SN5&3&3\\ 
SP1&4&4\\ 
SP2&6&5\\ 
SP3&3&3\\ 
SP4&4&5\\ 
SP5&2&3\\ 
\bottomrule
\end{tabular}
\end{adjustbox}
\caption{Participants' Expectation and Experience of the Task Difficulty}
\begin{tablenotes}
\vspace*{-0.4cm}
\item[1]\footnotesize{S=Spring, J=JSF, P=Priming, N=Non-Priming (See Table \ref{table:useCases} for labeling.)}
\end{tablenotes}
\label{table:expectations}
\end{table}
\small
\section{Survey}
\label{sec:survey}
\begin{itemize}
    \item \textit{Expectation} asked before solving the task:\\ What is your expectation? Overall, this task is?\\  \textbf{1: Very difficult - 7: Very easy}
    \item \textit{Experience} asked after solving the task:\\ Overall, this task was ...?  
    \textbf{1: Very difficult - 7: Very easy}
    \item I have a good understanding of security concepts. (Q1)\\
    \textbf{1: Strongly disagree - 7: Strongly agree} 
    \item How often do you ask for help facing security problems? (Q2)\\
    \textbf{1: Never - 7: Every time} 
    \item How often are you asked for help when somebody is facing security problems? (Q3)
    \textbf{1: Never - 7: Every time} 
    \item How often do you need to add security to the software you develop in general (Primed Group: apart from this study)? (Q4)\\
    \textbf{1: Never - 7: Every time}
    \item How often have you stored passwords in the software you have developed (Primed Group: apart from this study)?\\
    \textbf{1: Never - 7: Every time}
    \item How would you rate your background/knowledge with regard to secure password storage in a database?
    \textbf{1: Not knowledgeable at all - 7: Very knowledgeable} 
    \item Do you think that you stored the end-user passwords securely?
    \item The Web framework (Spring/JSF) supported me in storing the end-user password securely. \textbf{1: Strongly disagree - 7: Strongly agree}
    \item The Web framework (Spring/JSF) prevented me in storing the end-user password securely. \textbf{1: Strongly disagree - 7: Strongly agree}
\end{itemize}

\section{Semi-structured Interview}
\label{sec:interview}
\begin{itemize}
    \item Do you think that you have stored the end-user passwords securely
    \begin{itemize} 
    \item No
    \begin{itemize} 
         \item Why? 
         \item \textit{Non-Primed Group:} Were you aware that the task needed a secure solution?
         \item What would you do, if you needed to store the end-user password securely?
    \end{itemize} 
    \end{itemize}
\begin{itemize}
 \item Yes (If yes, further questions below)
\end{itemize}
 \item What is the purpose of hashing?
 
 \item What is the purpose of salting?

 \item Did you hash the end-user password?

 \item Did you salt the end-user password?

 \item Did the Web framework \textit{(Spring/JSF)} support you in storing the end-user password securely? Please explain your answer.
 \item Did the Web framework \textit{(Spring/JSF)} prevent you from storing the end-user password securely? Please explain your answer.

 \item Imagine you were working in a company and you had exactly the same task as you got today with exactly the same task description and time constraint. Would you have solved the task in exactly the same way as you solved it today?

 \item Did you solve the task functionally?

 \item Did you solve the task securely?

\end{itemize}
\subsection*{Further Interview Questions:} 
(If participants believe they have stored the end-user password securely.)
\begin{itemize}

\item \textit{Non-Primed Group:} How did you become aware of the necessity of security in this task? At which point did you decide to store the end-user password securely?

\item What did you do to store the user password securely?

\item Please name all steps taken in order to store the end-user password securely
\item Do you think your solution is optimal? Why or why not?

\item What were the general problems you encountered when implementing secure end-user password storage?

\item \textit{Primed Group:} Do you think you would have stored the end-user passwords securely if you had not been told about it? Please explain your answer.

\end{itemize}

\clearpage
\clearpage 

\section{Study Task for both Frameworks}
\label{sec:studyTask}
\fbox{\includegraphics[page=1,scale=.7]{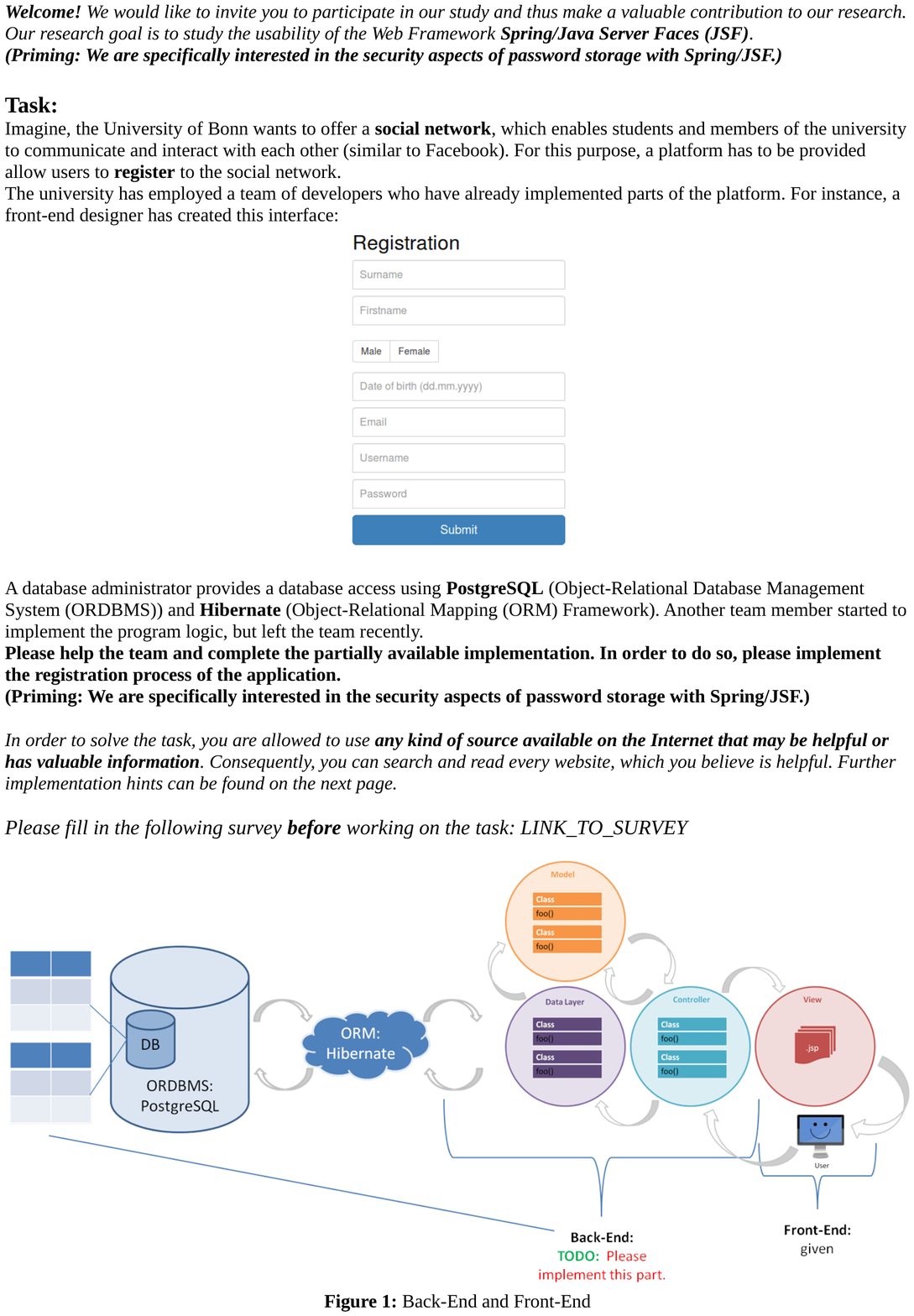}}

\clearpage

\section{Implementation hints for JSF}
\label{sec:jsfhints}
\fbox{\includegraphics[page=1,scale=0.77]{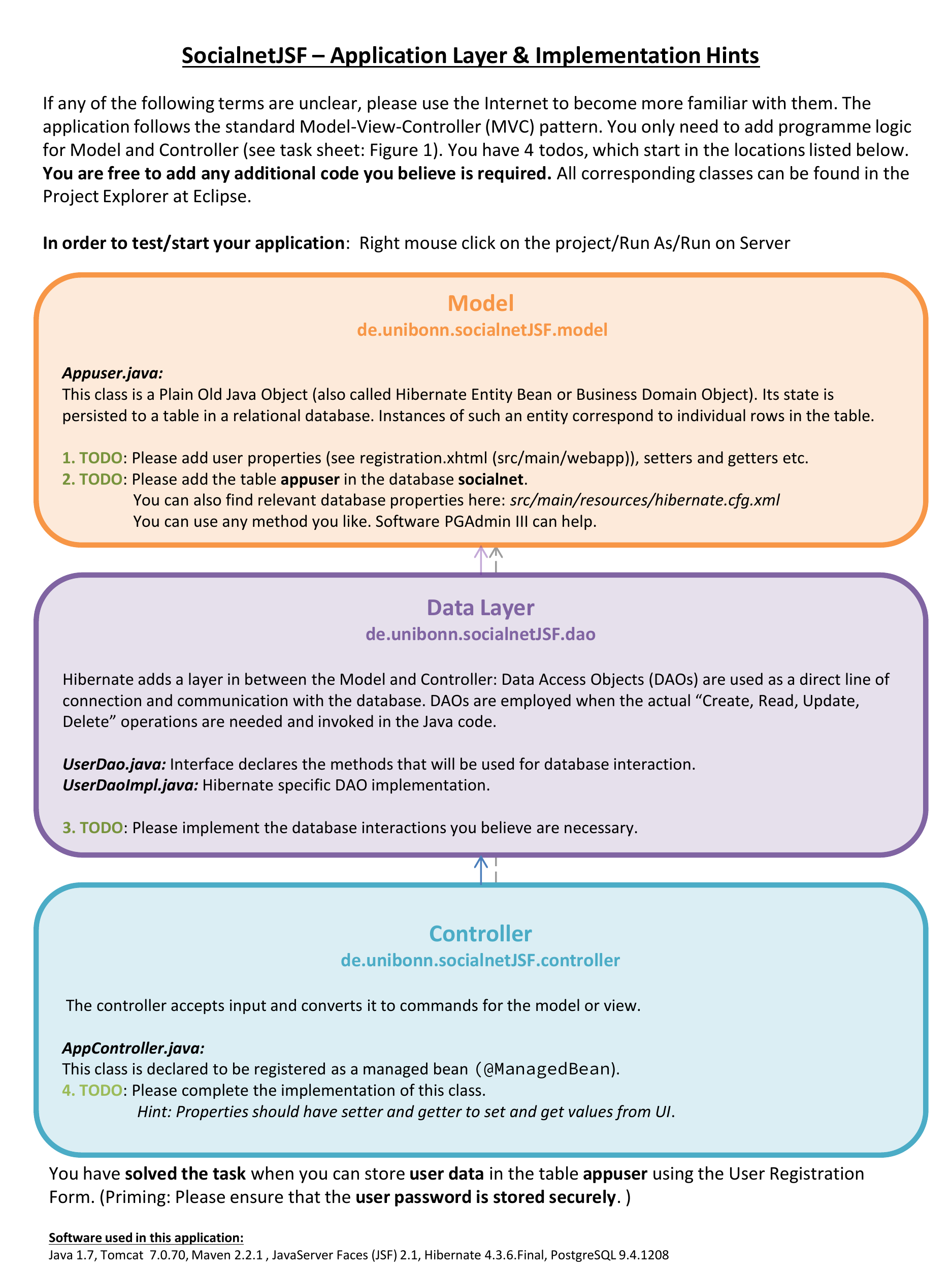}}

\clearpage

\section{Implementation hints for Spring}
\label{sec:springhints}
\fbox{\includegraphics[page=1,scale=0.77]{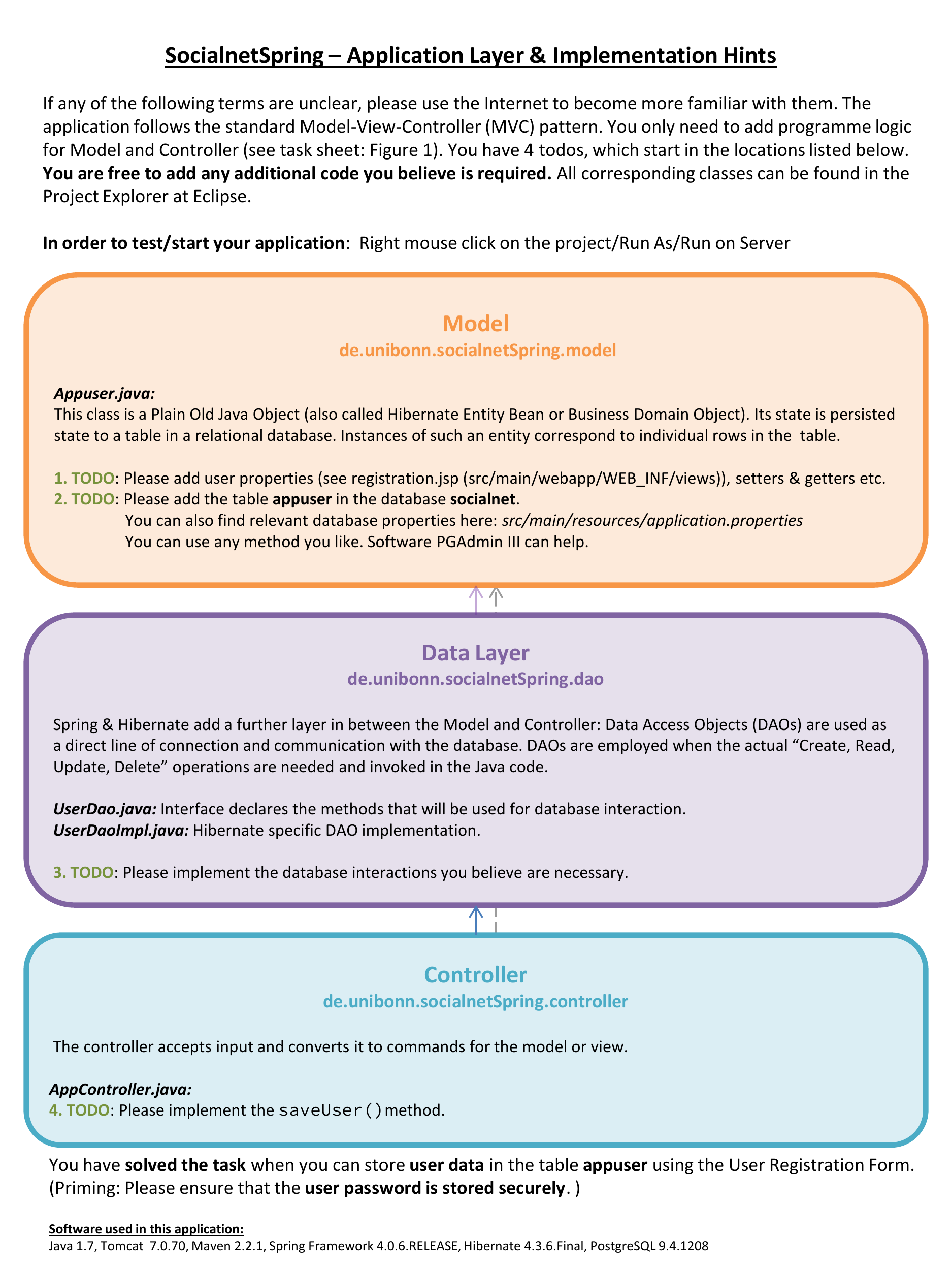}}


\end{appendix}

\end{document}